\documentclass[mathleft]{an}
\usepackage{graphicx}
\usepackage{times}
\begin{document}
\newcommand{\mbh}{M_{\bullet}}
\newcommand{\rg}{r_{\rm{}g}}
\newcommand{\rms}{r_{\rm{}ms}}

\tolerance=200
\setlength{\emergencystretch}{2em}

\Pagespan{001}{001}
\Yearpublication{2007}%
\Yearsubmission{2006}%
\Month{00}%
\Volume{000}%
\Issue{00}%

\title{Theoretical aspects of relativistic spectral features}

\author{Vladim\'{\i}r~Karas\thanks{\email{vladimir.karas@cuni.cz}\newline}}
\titlerunning{Theoretical aspects of relativistic spectral features}
\authorrunning{V.~Karas}
\institute{Astronomical Institute, Academy of Sciences, Bo\v{c}n\'{\i}~II, 
CZ-14131~Prague, Czech Republic}

\received{.. ... 2006}
\accepted{.. ... 2006}
\publonline{later}


\abstract{The inner parts of black-hole accretion discs shine in X-rays
which can be monitored and the observed spectra can be used to trace
strong gravitational fields in the place of emission and along paths of
light rays. This paper summarizes several aspects of how the spectral
features are influenced by relativistic effects. We focus our attention
onto variable and broad emission lines, origin of which can be attributed
to the presence of orbiting patterns -- spots and spiral waves in the
disc. We point out that the observed spectrum can determine parameters
of the central black hole provided the intrinsic local emissivity is
constrained by theoretical models.}

\maketitle

\section{Introduction}
During the last four decades a picture of galactic nuclei has emerged in
which massive black holes reside in the centre. In active galactic
nuclei (AGN) accretion discs form around the core and provide most of
the radiation emerging from that region. According to this scheme,
intense X-rays originate just a few gravitational radii from the black
hole horizon. Manifestation of strong gravity has been searched in X-ray
spectra of AGN with the main aim of finding the firm evidence
for the central black hole (BH) and determining its parameters, namely,
mass and angular momentum (for a detailed exposition of the subject, see
Peterson 1997; Krolik 1999).

Broad spectral features have been expected in X-rays on the basis
of the model in which the iron line is formed on the surface of a
geometrically thin, optically thick and relatively cold medium after
irradiation by a primary source (Fabian et al.\ 1989). Nowadays, the
general relativistic (GR) iron line profiles provide a powerful tool to
measure the mass of the black hole in AGN as well as Galactic black-hole
candidates. Blandford \& McKee (1982) and Stella (1990) have
developed the reverberation technique that employs a response of the
line profile following variations of the illuminating primary source.
Along the same line of thought, Matt \& Perola (1992) proposed to employ
variations of the time-integrated line properties -- equivalent width,
centroid energy and the line width. The method was developed further by
many authors (see Fabian et al.\ 2000; Reynolds \& Nowak 2003 for
reviews and references). Apart from the mass $\mbh$, the main parameters
of these models are the functional dependency of the intrinsic
emissivity over the disc surface, $I(r)$, the specific angular momentum $a$
of the black hole and the inclination $\theta_{\mathrm{o}}$ of the
accretion disc with respect to observer's line of sight. High resolution 
in spectral and time domains is crucial to accomplish the analysis.

This paper concentrates on gravitational effects that act on the
spectral features by smearing them and moving them among energy bins. 
In this way gravity exerts the influence on the ultimate form of the 
observed spectrum. We start with a brief summary of the equations
describing intensity (and polarization) propagation in the field of 
a black hole. Detailed exploration of broad spectral features can be 
considered as a method complementary to studying the overall X-ray continuum (Shaffee
et al.\ 2006) on one side and narrow spectral features (Turner et al.\
2004) on the other extreme. 

Relativistic effects are often discussed in
terms of geometrical optics with photons travelling through the empty
spacetime -- this formulation is suited quite well to various flavours
of `hot spots' on the disc surface, but plasma exists also above the
disc where it influences the observed signal to a certain degree. 
Theoretical approaches have been developed that can tackle more complicated
situations, such as the case of dispersive media which may be able to
address, more accurately, simultaneous observations in mutually remote
parts of the electromagnetic spectrum. 

In the second part of the paper we discuss flares and spots as a model
for X-ray variability: multiple spots are created on the surface of an
accretion disc following the intense irradiation. The observed signal is
then modulated by relativistic effects. This scheme is
testable and it captures many properties of present observations. 

Mean spectra of orbiting spots resemble those of axisymmetric rings and
we briefly discuss this degeneracy. It is inherent to time integrated
data and can be  resolved by increasing sensitivity and time resolution
of the observations.

In the present context, the spots represent any kind of localized
non-axisymmetric features residing on the disc surface, sharing its
orbital motion and modulating the signal. They are thought to result
from the illumination of a relatively cold gas of the disc after
magnetic reconnection events. Such a possibility was proposed for the
X-ray emitting region by Galeev et al.\ (1979) and developed in many
papers (Poutanen \& Fabian 1999; Merloni \& Fabian 2001;
\.Zycki 2002; Czerny et al.\ 2004, and references cited therein). 

The flare/spot scenario can be considered as a generalized model of the
phenomenological lamp-post geometry where a point-like X-ray source is
located at a specified height on the disc axis (Henri \& Pelletier 1991;
Martocchia \& Matt 1996). Martocchia et al.\ (2000) and Dov\v{c}iak et
al.\ (2004b) demonstrated by confidence contour analysis in the Kerr
metric that this scheme can be used to determine free parameters
{\em{}if} the resolution of our observations is sufficient and physical
mechanisms underlying the line features and the continuum are modeled
with sufficient precision, based on a sound understanding of the
underlying physics. 

New steps towards a detailed physical analysis of 
the radiation mechanisms forming X-ray spectra have been pursued by
several groups (Ballantyne et al.\ 2001; Nayakshin et al.\ 2000; Done 
\& Nayakshin 2001; R\'o\.za\'nska et al.\ 2002; Collin et al.\ 2004;
Kallman et al.\ 2004; Ross \& Fabian 2005) and adapted
specifically for the orbiting spot model (Czerny et al.\ 2004; Goosmann
et al.\ 2006). A similar line of reasoning stressing that the intrinsic
emission of the disc medium has to be accurately modelled has been now
adopted also by Brenneman \& Reynolds (2006). Although the implicit
assumptions of the codes and numerical approaches  of different authors
are quite divers, useful comparisons were performed  within the
overlapping range of the parameter space and can be found in the
above-mentioned papers.

More complex geometries of the emitting region should be also explored,
such as spiral waves propagating across the disc. This comes in accord
with recognition of the importance magnetic torques may have (Krolik et
al.\ 2005), but we can safely conclude that strong gravity of the
central BH is very likely the main agent shaping the overall form of the
X-ray spectral features from the inner disc.

\section{Preliminaries}
AGN variability time-scales extend down to a fraction of an hour and 
even less; this is comparable with the Keplerian orbital period near a
massive BH. The period of matter revolving along $r=\mbox{const}$
circular trajectory is $t_{\rm{}orb}=310(r^\frac{3}{2}+a)M_7$ seconds as
measured by a distant observer, where $M_7$ is the BH mass in units of
$10^7$ solar masses.\footnote{For the gravitational field we assume the
Kerr spacetime which is described by metric (\ref{eq:metric}) below in
the text. Lengths are expressed in units of the gravitational radius,
$\rg{\equiv}G{\mbh}/c^2{\doteq}1.48\times10^{12}M_7$~cm. The
dimension-less angular momentum $a$ adopts values in the range
$-1\leq{a}\leq1$. Positive values correspond to co-rotating motion, 
while negative values describe counter-rotation (many papers assume that
the accretion disc co-rotates, although such an assumption may not be
necessarily true). Circular orbits of free particles are possible above
the marginally stable orbit $r=\rms(a)$ for the corresponding angular
momentum (Bardeen et al.\ 1972).}

The orbital period $t_{\rm{}orb}$ is not much longer than the
light-crossing time, $t_{\rm{c}}$, because at distances of the order of
a few $\rg$ the bulk speed of accreted material is comparable with the
speed of light. Other time-scales relevant for black-hole accretion
discs -- thermal, sound-crossing, and viscous -- are typically longer
than $t_{\rm{}c}$. Because gravitation governs the motion near the
horizon, characteristic time-scales of the observed variability can be
scaled with $\mbh$. There are, however, various subtleties; the light
travel time obviously depends on the inclination of the system with
respect to an observer and also on scattering events that photons may
experience in the disc corona.

The reprocessed radiation reaches the observer from different regions of
the system. Furthermore, if strong-gravity plays a crucial role photons may even
follow multiple separate paths, joining each other at the observer. Individual
rays experience unequal time lags -- for purely geometrical reasons and
for relativistic time dilation. The final signal is smeared by GR
effects that become dominant in the close vicinity of the hole and are
relevant for the source spectrum and its variability (Laor 1991).

The irradiation of the disc from a fluctuating source of primary X-rays
does not lead to instantaneous response; time delays occur depending on
the geometrical and physical state of the gaseous material (Blandford
and McKee 1982; Taylor 1996). A single variability event is therefore
made of a primary flare and a complex response due to reflection by the
disc matter at various distances from the hole. Here we will neglect the
possibility of non-gravitational delays, neither we will touch an 
interesting possibility that a part of reprocessed radiation is
actually due to fast particles impinging on the disc medium (Ballantyne
\& Fabian 2003; Antonicci \& Gomez de Castro 2005). Instead we will
concentrate ourselves on geometrical effects connected with GR. Time-resolved 
spectroscopy offers the most accurate and practical method of revealing 
GR effects which is accessible with present-day technology. 

In the case of persisting patterns ($t\gg{}t_{\rm{c}}$) a substantial
contribution to the variability is caused by the orbital motion
(Abramowicz et al.\ 1991; Mangalam \& Wiita 1993). Two types of such
patterns have been examined in more detail, assuming the disc obeys a
strictly planar geometry: spots (which could be identified with vortices
in gaseous discs; cf.\ Abramowicz et al.\ 1992; Adams, Watkins 1995;
Karas 1997), and spiral waves (Tagger et al.\ 1990; Sanbuichi et al.\
1994). Though it may not be feasible with current technology, these
basic possibilities are distinguishable. Namely, the difference between
short-living flares (which die on the dynamical time-scale) in contrast
to the enduring features can be identified in time-resolved spectra
(Lawrence \& Papadakis 1993; Kawaguchi et al.\ 2000), but in order to
resolve it the geometry of the system has to be constrained. This is not 
an easy task because obscuration e.g.\ by a warped disc or off-equatorial 
clouds may complicate the analysis (Fukue 1987; Abrassart \& Czerny 2000; 
Karas et al.\ 1992, 2000; Hartnoll \& Blackman 2000). 

\begin{figure*}[tb]
\hfill~
\includegraphics[height=0.34\textwidth]{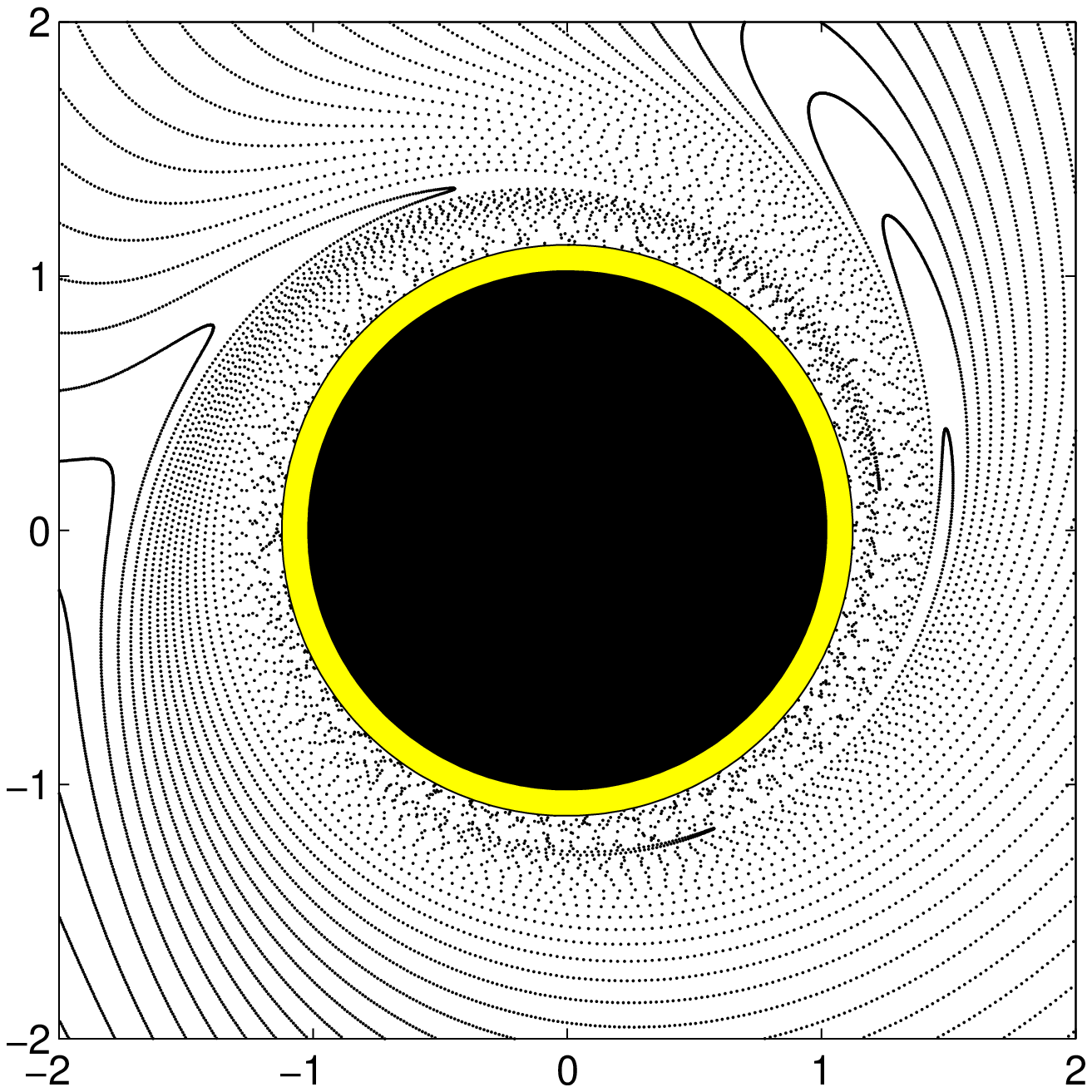}
\hfill~
\includegraphics[height=0.34\textwidth]{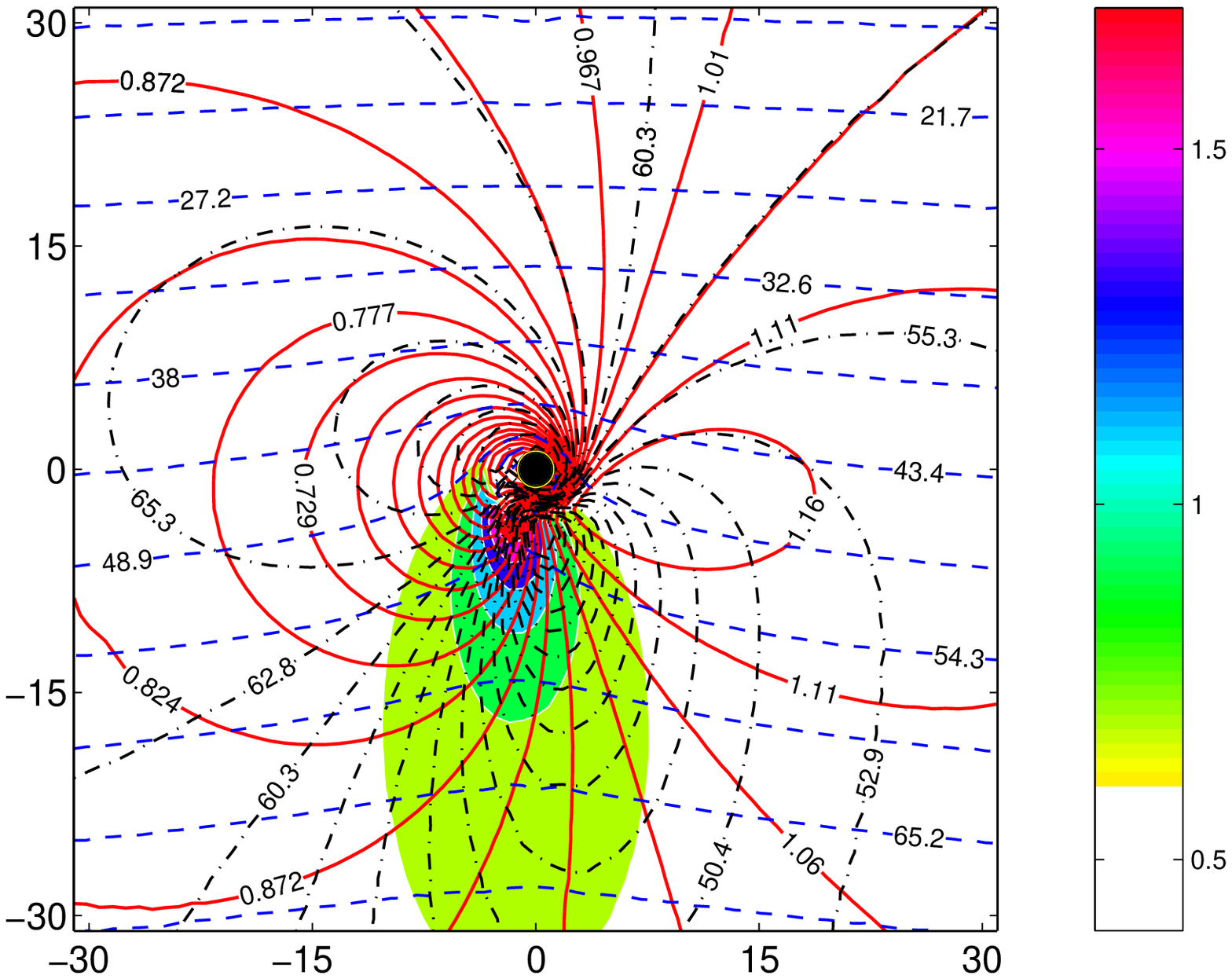}
\hfill~
\caption{Describing 
the GR effects acting on radiation from a source near a Kerr black hole
($a=0.998$). Photons emerge from different points of the equatorial
plane and proceed to a distant observer (located towards top of the
figure), where they are collected in a detector. The light experiences
an energy change and focusing effects which influence the observed
spectrum. Time delays between photons are important for time-resolved
spectroscopy. Left panel: the relativistic frame-dragging is illustrated
here by plotting the points of intersection of the rays emerging from
the equatorial plane. A polar grid was used, uniformly spaced in the
observer plane (inclination $\theta_{\mathrm{o}}=60$~deg;
$\theta_{\mathrm{o}}=90$~deg corresponds to edge-on observation, i.e.\
in the plain of the disc). The points of intersection form ellipses
which exhibit progressive distortion near the hole.  The frame-dragging
is clearly visible (the hole rotates counter-clockwise; the radii of
horizon and of the co-rotating innermost stable orbit are indicated by
circles around the origin). Right panel: Four sets of contours are
shown. They correspond to (i)~redshift function (solid lines); (ii)~time
delay (dashed); (iii)~emission angle with respect to the local normal
direction (dot-dashed); and (iv)~radiation flux enhancement due to
gravitational lensing (different levels of shading). Knowledge of these
quantities is sufficient to compute the GR effects that are expected in
observed spectra. Both plots have been constructed in Boyer-Lindquist
coordinates; lengths are expressed in units of $G\mbh/c^2$.}
\label{fig:del}
\end{figure*}

\section{GR effects on light from accretion discs}
\begin{figure}[tb]
\includegraphics[width=0.46\textwidth]{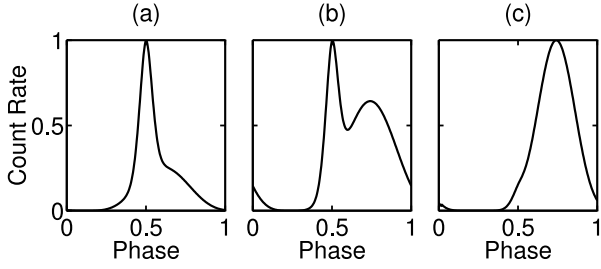}
\caption{Typical profiles of photometric light-curves from an orbiting 
spot (Karas 1996). The flux has been normalized to the maximum value.
Parameters of each panel: (a)~$r=3$, $\theta_{\mathrm{o}}=80$~deg
(lensing peak dominates the lightcurve near the orbital phase $0.5$);
(b)~$r=44$, $\theta_{\mathrm{o}}=80$~deg; (c)~$r=44$,
$\theta_{\mathrm{o}}=20$~deg (Doppler peak dominates near the phase
$0.75$). Each curve has been normalized to the maximum flux. The orbital
phase captures one revolution of the spot near a maximally rotating black hole.}
\label{fig:fitformula}
\end{figure}

The role of the central gravitational field is illustrated in
figure~\ref{fig:del}. Here, a view of the disc plane is shown as seen
from above, along the BH rotation axis (it is assumed that the disc
co-rotates with the hole around their common axis). What we see is a
complicated interplay of geometrical effects originating from the high
curvature and rotation of the BH spacetime, and the abberation effects
which result from the orbital motion of matter. All these need to be 
taken into account when determining the expected lightcurves 
(figure~\ref{fig:fitformula}). Proceeed to the contribution by G.~Matt 
in this volume for more details on GR effects seen in X-ray line spectra .

The geometrical optics approximation is adequate and hence the task is
reduced to integration of null geodesics. Many people have tackled this
problem by following Synge (1967) and assuming a fixed geometry of the
spacetime. Such an approach is well substantiated because it is the
black hole that is of particular interest, while the disc gravity plays
only a secondary role on the spacetime metric (it may be important,
however, in gamma-ray burst models in which a companion neutron star 
becomes tidally disrupted). This situation is useful also for
purposes of illustration (the case of a heavy disc poses a
mathematically and computationally more challenging task because it
requires to search for a simultaneous solution of Einstein's equations;
see Karas et al.\ 1995; Usui et al.\ 1998).

The basic properties of BH X-ray spectra from the innermost regions of
an accretion disc are well-known (Fabian et al.\ 2000; Reynolds \& Nowak
2003). The so called disc--line scheme has been widely applied because
it allows us to study how the spectral characteristics are formed when
accretion proceeds in strong gravity (Fabian et al.\ 1989). The main
spectral components are the X-ray continuum -- primary and reflected,
and the lines -- notably, the iron K$\alpha$ line (a doublet with the
rest energy of $\simeq6.4$~keV) and the higher ionisation lines
($\simeq6.7$--$6.9$~keV). The line is likely formed within a 
narrow interval of radii outside $\rms$ with the local intensity $I(r)$
decreasing outwards. However, the effects of the BH gravity are smeared
in real observations in which the signal comes from different,
insufficiently resolved regions. Because of the integration time, the
mean spectrum does not distinguish an orbiting source, such as a spot,
from the entire ring.

The gravitational field is described by the metric of Kerr black hole 
(Misner et al.\ 1973):
\begin{eqnarray}
ds^2&=&-\frac{\Delta}{\Sigma}\Big(dt-a\sin^2\theta\;d\phi\Big)^2
+\frac{\Sigma}{\Delta}\;dr^2+\Sigma\;d\theta^2
\nonumber \\
&&+\frac{\sin^2\theta}{\Sigma}\Big[a\;dt-\big(r^2+a^2\big)\;d\phi\Big]^2
\label{eq:metric}
\end{eqnarray}
in Boyer-Lindquist spheroidal coordinates, where functions $\Delta(r)$
and $\Sigma(r,\theta)$ are known. Self-gravitation of the accreted gas
is not taken into account in our discussion here. Among the properties
of the metric (\ref{eq:metric}) is the fact that the horizon occurs at 
$\Delta(r)=0$. Assuming sub-maximal rotation of the BH, $|a|\leq1$, the
outer radius is found at the dimension-less $r=1+(1-a^2)^{1/2}$ and it
hides the singularity from a distant observer. As a purely GR effect,
once $a\neq0$ all particles and photons are dragged to co-rotation with
the black hole.

Neglecting the disc gravity is an entirely adequate assumption for the
inner regions of AGN accretion discs (Novikov et al.\ 1973). However,
the `standard' black hole disc model does not seem to reproduce the
required broadening of the iron line even if it is supplemented by a
local corona above the disc. Generalized approaches have been discussed
and the standard picture of the disc continuum spectrum was reconsidered
to account for self-gravity (Laor \& Netzer 1989). Karas et al.\ (1995)
examined the gravitational effects the disc might have on the observed
line profiles; the increase of the equivalent width was found to be only
moderate.

The very existence of the innermost stable orbit is due to general
relativity. For a co-rotating equatorial disc this orbit has the radius
(Bardeen et al.\ 1972)
\begin{equation}
\rms = 3+z_2-\big[\left(3-z_1)(3+z_1+2z_2\right)\big]^\frac{1}{2},
\label{velocity1}
\end{equation}
where $z_1 = 1+\alpha_+\alpha_-[\alpha_++\alpha_-]$,
$\alpha_{\pm}=(1{\pm}a)^\frac{1}{3}$, and $z_2 =
(3a^2+Z_1^2)^\frac{1}{2}$. The marginally stable radius spans the range
from $\rms=1$ (for $a=1$, i.e.\ a maximally co-rotating BH) to $\rms=6$
(for $a=0$, a static BH). Black-hole rotation is limited to an
equilibrium value by photon recapture from the disc (which for the
standard disc model yields an equilibrium value of $a\dot{=}0.998$;
Thorne 1974) and by magnetic torques (Krolik et al.\ 2005).

For a source of light obeying purely prograde Keplerian motion the
orbital velocity is
\begin{equation}
 v^{(\phi)}=\Delta^{-1/2}\left(r^2-2ar^{1/2}+{a}^2\right)\left(r^{3/2}
  +{a}\right)^{-1}.
\end{equation}
Here, velocity is defined with respect to a locally non-rotating
observer at the corresponding radius. The corresponding angular velocity
is $\Omega(r;a)=(r^{3/2}+{a})^{-1}$, which also determines the orbital 
period $t_{\rm{}orb}$. In order to derive the time and frequency  measured
by a distant observer, one needs to take into account the Lorentz factor
associated with the orbital motion,
\begin{equation}
 \Gamma = \frac{\left(r^{3/2}+{a}\right)\Delta^{1/2}}{r^{1/4}\;
 \sqrt{r^{3/2}-3r^{1/2}+2{a}}\;\sqrt{r^3+{a}^2r+2{a}^2}}\,.
\end{equation}

By ignoring all other sources of gravitation except the central black
hole we assume there are no secondary bodies influencing 
gravitationally the inner disc. (This may not be true in case of binary
BHs or in the presence of global magnetic fields that can warp the disc
away from a unique plane.) The radiation field is then determined by
solving Maxwell's equations for the electromagnetic field tensor and its
dual in a fixed curved spacetime. In the vacuum we write:
${F^{\mu\nu}}_{\!;\nu}=0$, and ${^\star F^{\mu\nu}}_{\!;\nu}=0$,
where the asterisk denotes a dual tensor. Electric field components can
be  obtained by projection onto  observer's four-velocity,
$E^\alpha=F^{\alpha\beta}u_\beta$, and an electromagnetic wave is 
defined as an approximate test-field solution of the form
\begin{equation}
F_{\alpha\beta}=\Re e\left[u_{\alpha\beta}\;e^{\Im \psi(x)}\right].
\end{equation}
Two scales are introduced at this point. The phase $\psi(x)$ is assumed
to be a rapidly varying function, while the amplitudes $u_{\alpha\beta}$
vary slowly. This allows us to define a wave vector, $k_\alpha\equiv
\psi_{,\alpha}$, which is parallelly transported along the null geodesics,
$k_{\alpha;\beta}\,k^\beta=0$, $k_\alpha k^\alpha=0$. The propagation
law in the empty space is (Anile \& Breuer 1974)
\begin{equation}
DF_{\alpha\beta}-2\theta F_{\alpha\beta}=0,
\end{equation}
where $\theta\equiv-\frac{1}{2}{k^\alpha}_{;\alpha}$ describes the
expansion of null congruences, $D\,\equiv\,u^\alpha\nabla_\alpha$.

\begin{figure*}[tb]
\hfill~
\includegraphics[width=0.3\textwidth]{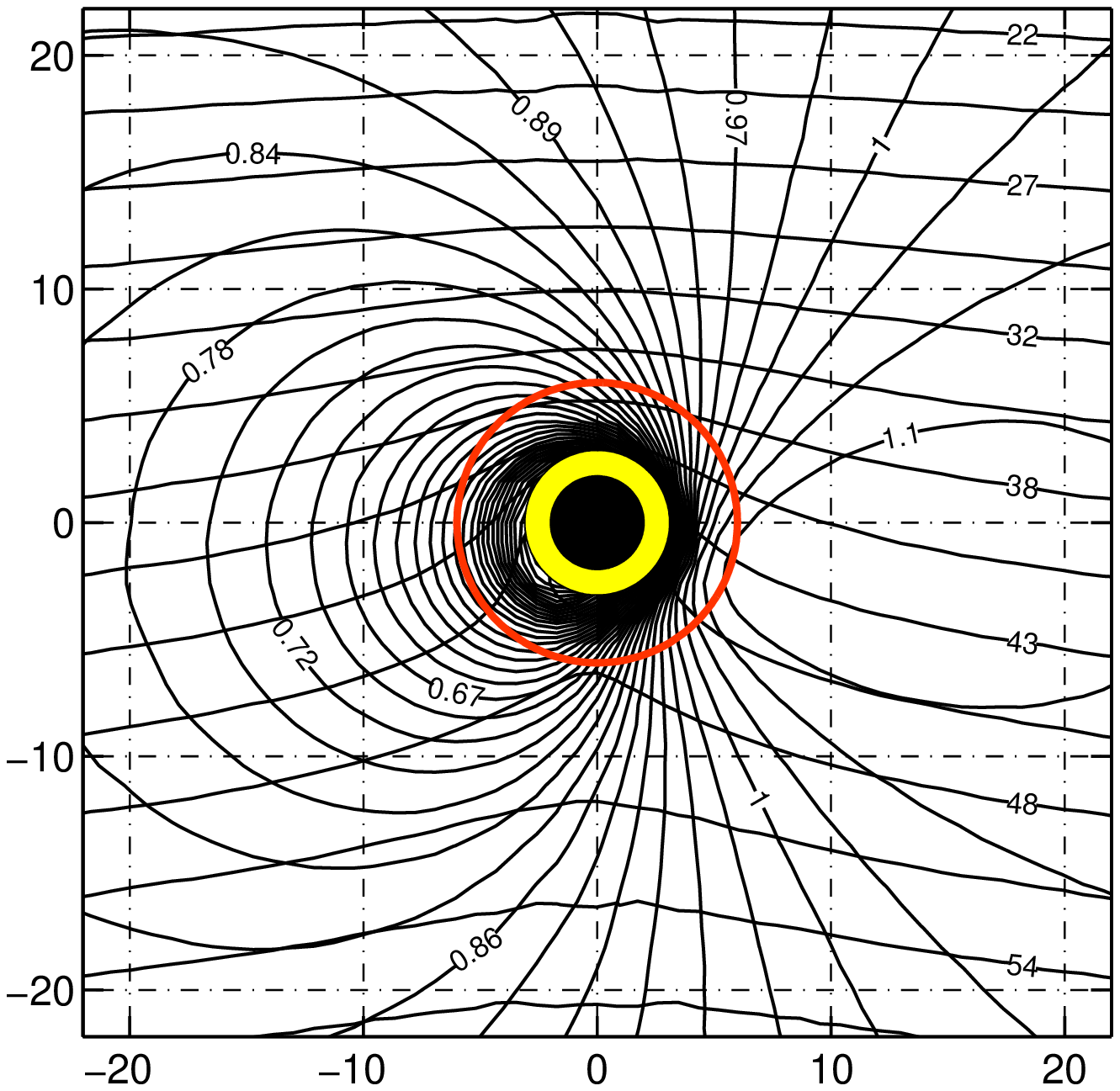}
\hfill~
\includegraphics[width=0.3\textwidth]{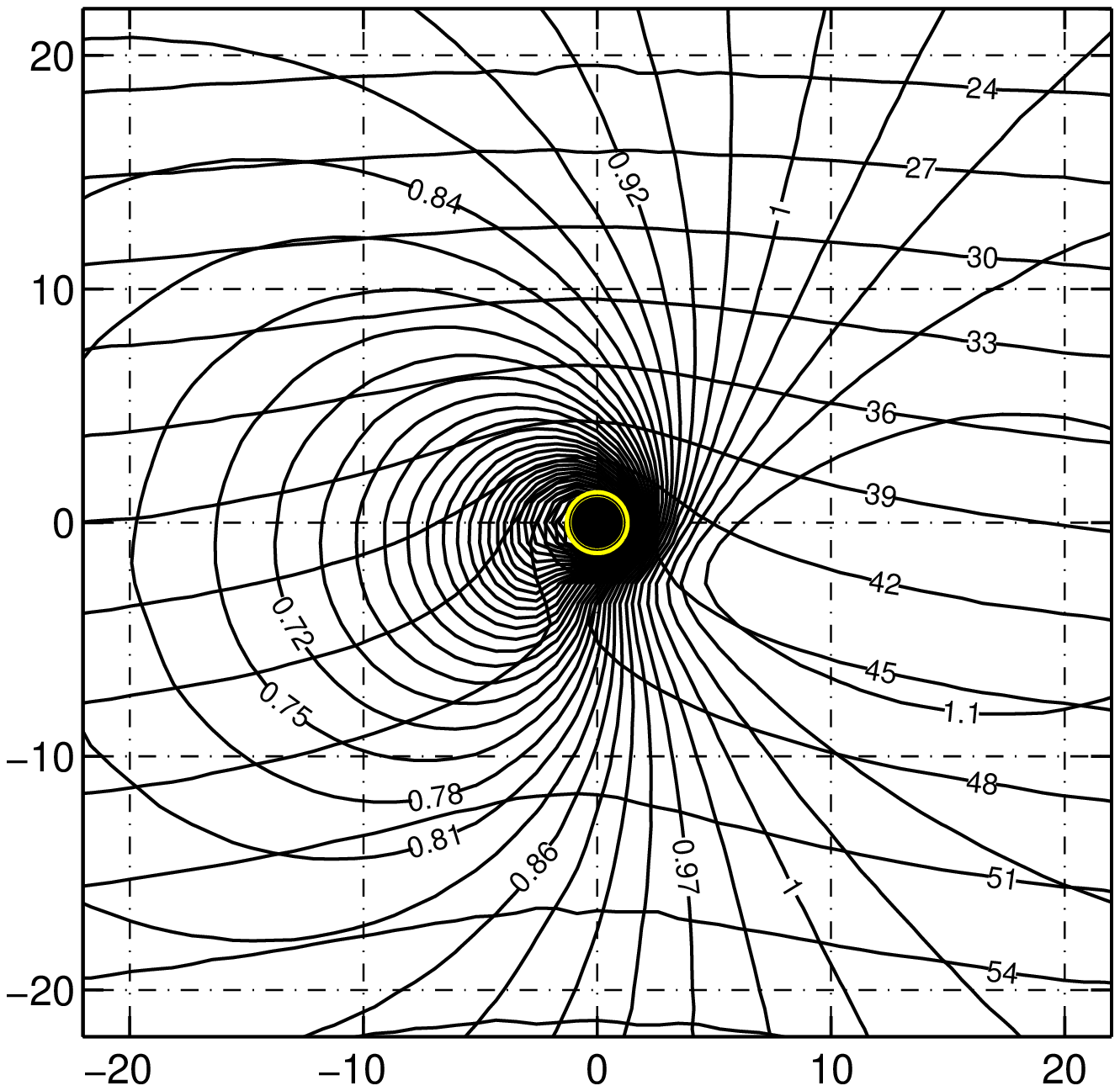}
\hfill~
\caption{Isocontours
of constant light-travel time $\delta{t}$ (approximately horizontal
lines) and of the redshift function $1+z$ (loops) are shown in the
equatorial plane of the Kerr BH. Two panels are shown for a non-rotating
BH ($a=0$, left), and for a maximally rotating BH ($a=1$, right). For
$a=0$, three circles are plotted with radii of the horizon, the circular
photon orbit, and the marginally stable orbit. The three radii appear to
coincide with each other in the maximally rotating case (co-rotating orbits are
assumed). In the plunging region (shaded) the form of the isocontours
depends on the assumed motion of the material, which no longer maintains
Keplerian $r=\mbox{const}$ orbits; magnetic fields likely play a major
role here. All values are given in geometrized units, observer's
inclination is $\theta_{\mathrm{o}}=50$~deg.}
\label{fig:del2}
\end{figure*}

Having in mind the applications to present-day X-ray observations, the
energy shifts, gravitational lensing and time delays are the principal
effects which originate from general relativity and can be tested with
data that are in our disposal. Polarimetric information goes beyond this
limit and has a capability of constraining parameters of our models at a
higher level. Different approaches to polarimetry and specific  issues
of X-ray polarimetry were discussed by various  authors, namely, Cocke
\& Holm (1972), Anile et al.\ (1974, 1977), Madore  (1974), Bi\v{c}\'ak
\& Hadrava (1975). In strong gravity, the covariant  definition of basic
polarimetric quantities is appropriate and it was developed in various
flavours: see Breuer \& Ehlers (1980) and  Anile (1989) for the
definition of the polarization tensor,
$J_{\alpha\beta\gamma\delta}\,\equiv\,\frac{1}{2}\langle 
F_{\alpha\beta}F_{\gamma\delta}\rangle$. A recent discussion was given
in Broderick \& Blandford (2003).

At first, projections
$J_{\alpha\beta}=J_{\alpha\beta\gamma\delta}\,u^{\gamma}\,u^{\delta}
=\langle E_\alpha \bar{E}_\beta\rangle$ are introduced. Four parameters
$S_A$ are then given by $S_{\!_A}\equiv \frac{1}{2}(k_\alpha
u^\alpha)^2F_{\!_A}$, where $F_{\!_A}$ $(A=0,\,\ldots\,3)$ are again
constructed by projecting the polarization tensor  (Anile \& Breuer
1974). These quantities satisfy the relations 
$J_{\alpha\beta}u^\beta=0$, $J_{\alpha\beta}k^\beta=0$, 
$\omega=u_\alpha k^\alpha$ and can be connected with the traditional 
definition of the Stokes parameters (Stokes 1852; Chandrasekhar 1960).

The normalized Stokes parameters are $s_1\,\equiv\,S_1/S_0$,
$s_2\,\equiv\,S_2/S_0$, and $s_3\,\equiv\,S_3/S_0$. The degree of linear
and of circular polarization is  $\Pi_l=\sqrt{s_1^2+s_2^2}$,
$\Pi_c=|s_3|$, and the total degree of polarization
$\Pi=\sqrt{\Pi_l^2+\Pi_c^2}$.

Upon propagation through an arbitrary (curved but empty)  space-time,
the radiation flux obeys the well-known relation, $F_{\!_{A,\,\rm
em}}\,dS_{_{\rm em}}=F_{\!_{A,\,\rm obs}}\,dS_{_{\rm obs}}$ from which
one can relate the redshift $z$ at the point of emission with the
redshift at a distant observer,
\begin{equation}
1+z=
\frac{(k_\alpha u_\alpha)_{\rm em}}{(k_\alpha u_\alpha)_{\rm obs}},\quad
S_{\!_A}=\frac{k_{\!_A}}{(1+z)^2dS}.
\label{eq:prop}
\end{equation}

In the Kerr metric--thin disc case, the redshift factor $z$ and the
local emission angle $\vartheta$ are given by
\begin{equation}
1+z=\frac{r^{3/2}-3r^{1/2}+2a}{r^{3/2}+{a}-\xi}, \quad
\cos\vartheta=\frac{g\eta^{1/2}}{r};
\end{equation}
$\xi$ and $\eta$ are constants of motion (they are connected with the
photon ray and exist in every axially symmetric and stationary
spacetime; see figure~\ref{fig:del2}). Let us remark that generalizations
to non-Keplerian motion of the reflecting material were discussed by
several authors: Reynolds \& Begelman (1997) discussed radiation from
the plunging region with matter in free-fall motion, and Dov\v{c}iak et
al.\ (2004) implemented this possibility in the {\sc{}ky} code. 
Also Fukue (2004) examined the non-negligible radial component of velocity. 

To explore GR effects from BH accretion discs, {\sc{}ky} is
currently the most versatile code available publicly and included in the
spectral fitting {\sc{}xspec} package. It allows its user to specify 
time-dependent/non-axisymmetric emissivity $I(t;r,\phi)$ of the disc (e.g.\ spiral
waves), explore the plunging region (e.g.\ falling blobs), and to vary
the black-hole angular momentum as well as the inner edge of the disc as
free parameters. The code has also capacity to study GR polarization;
this option cannot be currently  exploited in practice and we need to
wait for future satellites to perform  the job. However, it is
interesting to note that a complementary task can be partially
accomplished with the help of ground based near-infrared observations of the
Galaxy Center flares (see also Hollywood \& Melia 1997). Future 
X-ray satellite data will be essential to
study GR in the neighbourhood of accreting black holes with outstanding 
precision.

\begin{figure}[tb]
\begin{center}
\includegraphics[width=0.33\textwidth]{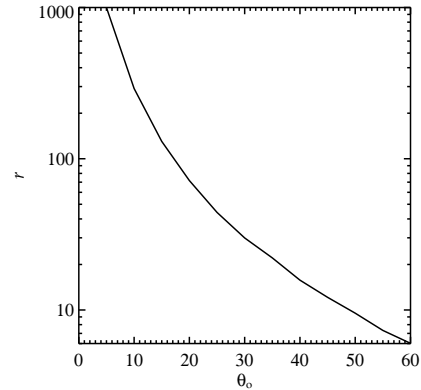}
\end{center}
\caption{Radius $r$ at which the maximum energy shift,
$g_+(r)_{|r=\mbox{const}}$, reaches its peak value as a 
function of the inclination $\theta_{\mathrm{o}}$. This shows 
how large excursions of the blue peak of the line can be 
expected. For $\theta_{\mathrm{o}}>60$~deg the maximum occurs 
at the marginally stable orbit where the inner edge of the disc
is often assumed (details in Pech\'a\v{c}ek  et al.\ 2005).
More complete information about the observed width of the line 
can be obtained from graphs of the redshift factor and the 
gravitational lensing, as they vary across the disc surface 
(see the paper by Matt in this volume).}
\label{fig:plotgm}
\end{figure}

In a typical situation of an AGN accretion disc, and especially for
small  radii and intermediate to large inclination angles, the line
emission comes from a small fraction of the orbit. This
implies that for observations with a moderate signal-to-noise ratios,
only a narrow blue horn is expected to appear for brief time. 
Figure~\ref{fig:plotgm} shows how large
variations of the line energy arise at different parts of
the orbit.

Eq.~(\ref{eq:prop}) shows that the polarization properties of the disc
emission are modified by the photon propagation in the gravitational
field and they may provide additional information about the field
structure. Since the reflecting medium has a disc-like geometry, a
substantial amount of linear polarization is expected because of Compton
scattering. The set of four Stokes parameters describes polarization
properties of the scattered light entirely. However, in order to compute
the observable characteristics, i.e.\ count rates of different
polarization modes, one has to combine the reflected component with the
primary continuum. The polarization degree of the resulting signal
depends on the mutual proportion of the two components.

The idea of using polarimetry to gain additional information about
compact objects is not a new one. In this context it was proposed by
Rees (1975) that polarized X-rays are of high relevance. Pozdnyakov et
al.\ (1979) studied spectral profiles of X-ray iron lines that result
from multiple Compton scattering (also Angel 1969; Bonometto et al.\
1970; Sunyaev \& Titarchuk 1985; Matt 1993). The possibility of 
detecting gravitational effects on polarized X-rays is very 
attractive as it has great potential in revealing black holes, but 
the idea still awaits precise formulation as well as the technology 
usable for its practical implementation. Temporal variations of
polarization were discussed, in particular the case of orbiting
spots (Pineault 1977; Connors et al.\ 1980; Bao et al.\ 1998). Further
aspects of GR were examined by Viironen \& Poutanen (2004), 
Dov\v{c}iak et al. (2004a), and Hor\'ak \& Karas (2006a,\,b), who 
studied the role of multiple images.

\section{Spots and spirals}
In their seminal papers, Cunningham \& Bardeen (1972, 1973) studied the
radiation from a star revolving in the equatorial plane of a maximally
rotating BH. They presented a detailed analysis of periodic variations
of the observed frequency-integrated flux that arrives at the observer.
The method turns out to be a very practical one, even three decades
later when computer capabilities have increased tremendously. However,
the potential of their idea cannot be fully exploited until the theory
determines the intrinsic emission that emerges locally from the disc.
Neither it is well-suited for timing analyses (current status of this 
effort is summarized by Goosmann in this volume).

Since first attempts, which were largely limited to photometry, many
people have developed various modifications of the original
semi-analytical approach, especially for the spectroscopy of accretion
discs: Cunningham (1975); Gerbal \& Pelat (1981); Asaoka (1989); Bao \&
Stuchl\'{\i}k (1989); Viergutz (1993); Rauch \& Blandford (1994);
Zakharov (1994); Bromley et al.\ (1997); Fanton et al.\ (1997); Bao et
al.\ (1998); Semer\'ak et al.\ (1999); Martocchia et al.\ (2000);
Schnittman (2005), and others. Computational challenges have been
largely overcome thanks to a combination of semi-analytical approaches
and sophisticated numerical algorithms. The analytical part of the work
is largely based on the remarkable property of the geodesic motion in
Kerr metric (due to Carter 1968), which is integrable and can be carried
out in terms of elliptical integrals (de Felice et al.\ 1974; Sharp
1979; \v{C}ade\v{z} \& Calvani 2005). The numerical approach often
employs pre-computed data that are stored for the subsequent light-ray
reconstruction (Karas et al.\ 1992; Dov\v{c}iak et al.\ 2004b; Beckwith
\& Done 2004). Whichever strategy is adopted, the ray-tracing core of the 
code has to be connected with a radiation transfer routine determining
the intrinsic spectrum and, indeed, the understanding of the transfer 
problem has been tremendously improved in recent years.

The idea of magnetic flares irradiating the disc surface (Merloni \&
Fabian 2001) gives a promising route towards a more complete physical
formulation of the `bright-spot' model that would be based on elementary
processes and help reducing the excessive number of degrees of freedom. 
By employing this approach, Czerny et al.\ (2004) computed the
reprocessed spectra and compared the predicted variability with
MCG--6-30-15 observational data. The actual size of the spot is linked
with the X-ray flux which originates in flares and depends on the
vertical height where the flares occur above the disc plane. The induced
rms variability is a function of the energy range of the observation and
the model parameters including the BH angular momentum and the position
of the disc inner edge. Goosmann et al.\ (2006) examined different cases
with the spot size and height ranging from a fraction of $\rg$ to
several $\rg$. It is interesting to notice that the lamp-post geometry
is a limiting case of the flare/spot model. These results can reproduce
the observed features, albeit the present data do not constrain all
parameters; for example the spot size cannot be determined.

\begin{figure*}[tb!]
\begin{center}
\includegraphics[width=0.66\textwidth]{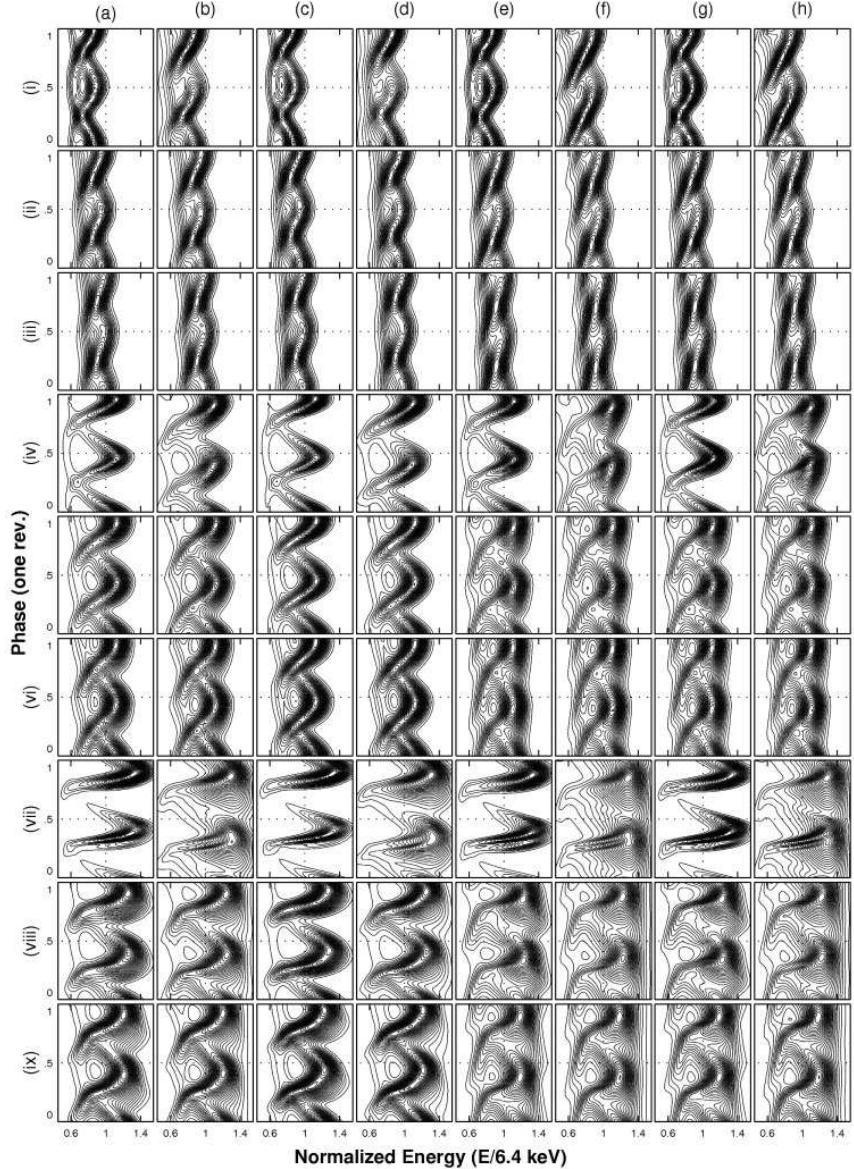}
\caption{An
examplary set of dynamical
spectra of a two-arm spiral pattern with a slowly decaying flare. Different
cases correspond to the form of the spiral, rotation parameter of the
BH, and observer's viewing angle. These translate to different amplitudes and
energy ranges spanned by observed spectra. The model parameters can be 
assorted in four categories: (A)~BH angular momentum: $a=0$ (columns denoted
by letters {\sf{a}}, {\sf{c}}, {\sf{e}}, {\sf{g}}), and $a=1$ ({\sf{b}},
{\sf{d}}, {\sf{f}}, {\sf{h}}).  (B)~Observer inclination:
$\theta_{\mathrm{o}}=20$~deg (the rows denoted by roman numbers
{\sf{i}}--{\sf{iii}}), $\theta_{\mathrm{o}}=50$~deg
({\sf{iv}}--{\sf{vi}}), and $\theta_{\mathrm{o}}=80$~deg
({\sf{vii}}--{\sf{ix}}).  (C)~The extent of the spiral; here,
the outer edge is located at: $r_0=7$ ({\sf{i}}, {\sf{iv}}, {\sf{vii}}),
$r_0=13$ ({\sf{ii}}, {\sf{v}}, {\sf{viii}}), and $r_0=20$ ({\sf{iii}},
{\sf{vi}}, {\sf{ix}}). (D)~The actual form
of the spiral structure is defined by the pitch angle $\arctan\alpha$
and the contrast from the background $\beta$: $\alpha=3$, $\beta=8$
({\sf{a}}, {\sf{b}}); $\alpha=\sqrt{3}$, $\beta=8$ ({\sf{c}}, {\sf{d}});
$\alpha=6$, $\beta=8$ ({\sf{e}}, {\sf{f}}); $\alpha=6$, $\beta=18$
({\sf{g}}, {\sf{h}}).  For details see Karas et al. (2001).}
\label{fig:prof}
\end{center}
\end{figure*}

A complementary picture assumes extended and evolving shapes of the
reflection area on the disc. Spiral shapes can arise from an extended
spot following its decay by shearing motion under various kinds  of
instabilities operating in the disc. Such patterns are transient but may
last over a substantial fraction of the orbital period, sufficiently
long to produce observable effects. Apart from AGN discs they have been
invoked also to explain the flares in Sagittarius~A$^{\star}$ (Tagger \&
Melia 2006) for which individual clumps rather than a continuous 
accretion flow seem to be responsible. Examples and useful templates of the
expected spectral line profiles were computed (Chakrabarti \& Wiita
1993; Karas et al.\ 2001; Hartnoll \& Blackman 2002; Fukumura \& Tsuruta
2004); see figure~\ref{fig:prof} where a combination of an instantaneous
flare and a persisting spiral was assumed. Here, the flare quickly
perishes and  the spiral dominates the spectrum. The spiral is described
by its pitch angle $\arctan\alpha$ (tightly wound spirals are described
by large values of the pitch angle)  and the contrast $\beta$
(well-defined patterns have a large  $\beta$). On the inner side the 
spiral terminates at $r=\rms$. In each panel the spectra were computed
in $10^2\times10^2$ energy--phase grid by our Kerr metric ray-tracing
code.

\section{Conclusions}
The influence of strong gravity on light provides a possibility to study
black holes. However, gravity is not the only agent that shapes the
observed spectra; emission mechanisms and radiation transfer in the
accreting gas need to be understood reliably. Remarkable progress has
been achieved on the way to model the physical processes which govern
the form of X-ray reflection spectra emerging from different regions of
accretion discs. In spite of this advance the current description still
contains some elements of the phenomenological approach and the
emission from non-axisymmetric patterns cannot be computed
entirely `from first principles'. Nonetheless, already at its present
formulation the model of a spotted disc helps us to constrain the
parameters of accreting black holes. The orbital radius of the peak
emission and the disc inclination can be inferred from the variations of
the observed line width and the centroid energy. $\mbh$ can
be estimated by comparing the measured orbital period with the value
expected for the derived radius. The model of spirals also gives the
possibility of tracing the surface structures on accretion discs. Once
these are determined the goal of measuring black hole rotation will be
feasible with unprecedented accuracy.

\acknowledgements
The author is very grateful to the Conference organizers for the
invitation. The financial support from the Academy of Sciences
(ref.~300030510) and the Center for Theoretical Astrophysics in Prague
is acknowledged.


\begin{thebibliography}{}
\vspace*{-1em}
\bibitem{10}Abramowicz M.~A., Bao G., Lanza A., Zhang X.-H.: 1991, A\&A 245, 454
\bibitem{20}Abramowicz M.~A., Lanza~A., Spiegel E.~A., Szus\-z\-kie\-wicz~E.: 1992, Nature, 356, 41
\bibitem{30}Adams F.~C., Watkins~R.: 1995, ApJ, 451, 314
\bibitem{40}Angel J.~R.~P.: 1969, MNRAS, 158, 219
\bibitem{50}Anile A.~M.: 1989, {\it Relativistic fluids and magneto-fluids} (Cambridge: Cambridge University Press)
\bibitem{52}Anile A.~M., Breuer R.~A: 1974, ApJ, 189, 39
\bibitem{60}Anile A.~M., Pantano~P.: 1977, Phys. Lett. A, 61, 215
\bibitem{70}Antonicci~A., G\'omez de Castro A.~I.: 2005, A\&A, 432, 443
\bibitem{80}Asaoka~I.: 1989, PASJ, 41, 763
\bibitem{90}Ballantyne D.~R., Fabian A.~C.: 2003, ApJ, 592, 1089
\bibitem{100}Ballantyne D.~R., Ross R.~R., Fabian A.~C.: 2001, MNRAS, 327, 10
\bibitem{110}Bao~G., Stuchl\'{\i}k~Z. 1992, ApJ, 400, 163
\bibitem{120}Bao~G., Wiita P.~J., Hadrava~P.: 1998, ApJ, 504, 58
\bibitem{130}Bardeen J.~M., Press W.~H., Teukolsky S.~A.: 1972, ApJ, 178, 347
\bibitem{140}Beckwith~K., Done~C.: 2004, MNRAS, 353, 362
\bibitem{150}Bi\v{c}\'ak~J., Hadrava~P.: 1975, A\&A, 44, 389
\bibitem{160}Blandford R.~D., McKee C.~F.: 1982, ApJ 255, 419
\bibitem{170}Bonometto~S., Cazzola~P., Saggion~A.: 1970, A\&A, 7, 292
\bibitem{175}Brenneman L.~W., Reynolds C.~S.: 2006, ApJ, in press
\bibitem{180}Breuer R.~A., Ehlers~J.: 1980, Proc. Roy. Soc. Lond. A, 370, 389
\bibitem{190}Broderick A.~E., Blandford R.~D.: 2003, MNRAS, 342, 1280
\bibitem{200}Bromley B.~C., Chen~K., Miller W.~A.: 1997, ApJ, 475, 57
\bibitem{210}\v{C}ade\v{z}~A., Calvani~M.: 2005, MNRAS, 363, 177
\bibitem{220}Carter~B.: 1968, Physical Review, 174, 1559
\bibitem{225}Chakrabarti S.~K., Wiita P.~J.: 1993, A\&A, 271, 216
\bibitem{230}Chandrasekhar~S.: 1960, {\em Radiative Transfer} (New York: Dover)
\bibitem{235}Cocke W.~J., Holm D.~A.: 1972, Nature Physical Science, 240, 161
\bibitem{237}Collin~S., Dumont A.-M., Godet~O.: 2004, A\&A, 419, 877
\bibitem{240}Connors P.~A., Stark R.~F., Piran~T.: 1980, ApJ, 235, 224
\bibitem{245}Cunningham C.~T.: 1975, ApJ, 202, 788
\bibitem{250}Cunningham C.~T., Baardeen J.~M.: 1972, ApJ, 173, L137
\bibitem{255}Cunningham C.~T., Baardeen J.~M.: 1973, ApJ, 183, 273
\bibitem{260}Czerny~B., R\'o\.za\'nska~A., Dov\v{c}iak~M., Karas~V., \& Dumont A.-M.: 2004, A\&A, 420, 1
\bibitem{265}de Felice~F., Nobili~L., Calvani~M.: 1974, A\&A, 30, 111
\bibitem{270}Done~C., Nayakshin~S.: 2001, MNRAS, 328, 616
\bibitem{280}Dov\v{c}iak~M., Karas~V., Matt~G.: 2004a, MNRAS, 355, 1005
\bibitem{290}Dov\v{c}iak~M., Karas~V., Yaqoob~T.: 2004b, ApJSS, 153, 205
\bibitem{300}Fabian A.~C., Iwasawa~K., Reynolds C.~S., Young A.~J.: 2000, PASP, 112, 1145
\bibitem{310}Fabian A.~C., Rees M.~J., Stella~L., White N.~E: 1989, MNRAS, 238, 729
\bibitem{320}Fanton~C., Calvani~M., de Felice~F., \v{C}ade\v{z}~A.: 1997, PASJ, 49, 159
\bibitem{325}Fukue~J.: 1987: Nature, 327, 600
\bibitem{330}Fukue~J.: 2004, Progress of Theor. Phys. Suppl., 155, 329
\bibitem{335}Fukumura~K, Tsuruta~S.: 2004, ApJ, 613, 700
\bibitem{340}Galeev A.~A., Rosner R., Vaiana G.~S.: 1979, ApJ, 229, 318
\bibitem{345}Gerbal~D., Pelat~D.: 1981, A\&A, 95, 18
\bibitem{350}Goosmann R., Czerny~B., Mouchet~M., Ponti~G., Dov\v{c}iak~M., Karas~V., R\'o\.za\'nska~A., Dumont A.-M.: 2006, A\&A, 454, 741
\bibitem{353}Hartnoll S.~A., Blackman E.~G.: 2000, MNRAS, 317, 880
\bibitem{355}Hartnoll S.~A., Blackman E.~G.: 2002, MNRAS, 332, L1
\bibitem{360}Henri~G., Pelletier~G.: 1991, ApJ, 383, L7
\bibitem{365}Hollywood J.~M., Melia~F.: 1997, ApJSS, 112, 423
\bibitem{370}Hor\'ak~J., Karas~V., 2006a, MNRAS, 365, 813
\bibitem{380}Hor\'ak~J., Karas~V., 2006b, PASJ, 58, 203
\bibitem{390}Kallman T.~R., Palmeri~P., Bautista M.~A., Mendoza~C., Krolik J.~H.: 2004, ApJSS, 155, 675
\bibitem{400}Karas~V.: 1996, ApJ, 470, 743
\bibitem{410}Karas~V.: 1997, MNRAS, 288, 12
\bibitem{415}Karas~V., Czerny~B., Abbrasart~A., Abramowicz M.~A.: 2000, MNRAS, 318, 547
\bibitem{420}Karas~V., Bao~G.: 1992, A\&A, 257, 531
\bibitem{430}Karas~V., Lanza~A., Vokrouhlick\'y~D.: 1995, ApJ, 440, 108
\bibitem{440}Karas~V., Martocchia~A., \v{S}ubr~L.: 2001, PASJ, 53, 189
\bibitem{450}Karas~V., Vokrouhlick\'y, Polnarev A.~G.: 1992, MNRAS, 259, 569
\bibitem{460}Kawaguchi~T., Mineshige~S., Machida~M., Matsumoto~R., Shibata~K.: 2000, PASJ, 52, L1
\bibitem{465}Krolik J.~H.: 1999, {\em Active Galactic Nuclei} (Princeton: Princeton University Press)
\bibitem{470}Krolik J.~H., Hawley J.~F., Hirose~S.: 2005, ApJ, 622, 1008
\bibitem{480}Laor~A.: 1991, ApJ, 376, 90
\bibitem{490}Laor~A., Netzer~H.: 1989, MNRAS, 238, 897
\bibitem{500}Lawrence~A., Papadakis~I.: 1993, ApJ, 414, 85
\bibitem{510}Madore~J.: 1974, Comm. Math. Phys., 38, 103
\bibitem{520}Mangalam A.~V., Wiita P.~J.: 1993, ApJ, 406, 420
\bibitem{540}Martocchia~A., Karas~V., Matt~G.: 2000, MNRAS, 312, 817
\bibitem{530}Martocchia~A., Matt~G.: 1996, MNRAS, 282, 53
\bibitem{550}Matt~G.: 1993, MNRAS, 260, 663
\bibitem{560}Matt~G., Perola G.~C.: 1992, MNRAS, 259, 433
\bibitem{570}Merloni~A., Fabian A.~C.: 2001, MNRAS, 328, 958
\bibitem{580}Misner C.~W., Thorne K.~S., Wheeler J.~A.: 1973, {\em Gravitation} (San Francisco: Freeman)
\bibitem{585}Nayakshin~S., Kazanas~D., Kallman T.~R.: 2000, ApJ, 537, 833
\bibitem{590}Novikov I.~D., Thorne K.~S.: 1973, in {\it Black Hole Astrophysics}, C.~DeWitt and B.\,S.\ DeWitt (eds), (New York: Gordon \& Breach), p.~{343}
\bibitem{600}Pech\'a\v{c}ek~T., Dov\v{c}iak~M., Karas~V., Matt~G.: 2005, A\&A, 441, 855
\bibitem{605}Peterson B.~M.: 1997, {\em Active Galactic Nuclei} (Cambridge: Cambridge University Press)
\bibitem{607}Pineault~S.: 1977, MNRAS, 179, 691
\bibitem{610}Poutanen~J., Fabian A.~C.: 1999, MNRAS, 306, 31
\bibitem{620}Pozdnyakov L.~A., Sobol I.~M., Sunyaev R.~A.; 1979, A\&A, 75, 214
\bibitem{630}Rauch K.~P., Blandford R.~D.: 1994, ApJ, 421, 46
\bibitem{640}Rees M.~J.: 1975, MNRAS, 171, 457
\bibitem{650}Reynolds C.~S., Begelman M.~C.: 1997, ApJ, 488, 109
\bibitem{660}Reynolds C.~S., Nowak M.~A.: 2003, Phys. Rep., 377, 389
\bibitem{665}Ross R.~R., Fabian A.~C.: 2005, MNRAS, 358, 211
\bibitem{670}R\'o\.za\'nska~A., Dumont A.-M., Czerny~B., Collin~S.: 2002, MNRAS, 332, 799
\bibitem{675}Sanbuichi~K., Fukue~J., Kojima~Y.: 1994, PASJ, 46, 605
\bibitem{680}Schnittman J.~D.: 2005, ApJ, 621, 940
\bibitem{690}Semer\'ak~O., Karas~V., de Felice~F.: 1999, PASJ, 51, 571
\bibitem{693}Shaffee~R., McClintock J.~E., Narayan~R., Davis S.~W., Li Li-Xin, Remillard R.~A.: 2006, ApJ, 636, L113
\bibitem{695}Sharp N.~A.: 1979, Gen. Rel. Grav., 10, 659
\bibitem{700}Stella~L.: 1990, Nature, 344, 747
\bibitem{710}Stokes~G.: 1852, Trans. Cambridge Phil. Soc., 9, 399
\bibitem{720}Sunyaev R.~A., Titarchuk L.~G.: 1985, A\&A, 143, 374
\bibitem{725}Synge J.~L.: 1967, MNRAS, 136, 195
\bibitem{730}Tagger~M., Henriksen R.~N., Sygnet J.~F., Pellat~R.: 1990, ApJ, 353, 654
\bibitem{735}Tagger~M., Melia~F.: 2006, ApJ, 636, L33
\bibitem{740}Taylor J.~A.: 1996, ApJ, 470, 269
\bibitem{750}Thorne K.~S.: 1974, ApJ, 191, 507
\bibitem{755}Turner T.~J., Mushotzky R.~F., Yaqoob~T., George I.~M., Snowden S.~L., Netzer~H. et al.: 2004, ApJ, 574, L123
\bibitem{760}Usui~F., Nishida~S., Eriguchi~Y.: 1998, MNRAS, 301, 721
\bibitem{770}Viergutz S.~U.: 1993, A\&A, 272, 355
\bibitem{775}Viironen~K., Poutanen~J.: 2004, A\&A, 426, 985
\bibitem{780}Zakharov A.~F.: 1994, MNRAS, 269, 283
\bibitem{790}\.Zycki P.~T.: 2002, MNRAS, 333, 800
\end{thebibliography}
\end{document}